\documentclass[12pt,preprint]{aastex}
\usepackage{natbib}
\shorttitle{Constraining X-ray Binary Jet Models}
\shortauthors{Markoff \& Nowak}

\begin{document}

\title{Constraining X-ray Binary Jet Models via Reflection}

\author{Sera Markoff\altaffilmark{1} and Michael A. Nowak}
\affil{Massachusetts Institute of Technology, Center for Space
Research, Rm. NE80-6035, Cambridge, MA 02139}
\email{sera,mnowak@space.mit.edu}

\altaffiltext{1}{NSF Astronomy \& Astrophysics
Postdoctoral Fellow}

\begin{abstract}
  
  Although thermal disk emission is suppressed or absent in the hard
  state of X-ray binaries, the presence of a cold, thin disk can be
  inferred from signatures of reprocessing in the $\sim2-50$ keV band.
  The strength of this signature is dependent on the source spectrum
  and flux impinging on the disk surface, and is thus very sensitive
  to the system geometry.  The general weakness of this feature in the
  hard state has been attributed to either a truncation of the thin
  disk, large ionization, or beaming of the corona region away from
  the disk with $\beta\sim0.3$.  This latter velocity is comparable to
  jet nozzle velocities, so we explore whether a jet can account for
  the observed reflection fractions.  It has been suggested that jets
  may contribute to the high-energy spectra of X-ray binaries, via
  either synchrotron from around $100-1000$ $r_{\rm g}$ along the jet
  axis or from inverse Compton (synchrotron self-Compton and/or
  external Compton) from near the base.  Here we calculate the
  reflection fraction from jet models wherein either synchrotron or
  Compton processes dominate the emission.  Using as a guide a data
  set for GX~339$-$4, where the reflection fraction previously has
  been estimated as $\sim10\%$, we study the results for a jet model.
  We find that the synchrotron case gives $< 2\%$ reflection, while a
  model with predominantly synchrotron self-Compton in the base gives
  $\sim 10-18\%$.  This shows for the first time that an X-ray binary
  jet is capable of significant reflection fractions, and that extreme
  values of the reflection may be used as a way of discerning the
  dominant contributions to the X-ray spectrum.

\end{abstract}


\section{Introduction}

X-ray binaries (XRBs) have been observed in several distinct states,
which are characterized by the relative strength of their soft and
hard X-ray emission components, as well as by their variability
properties \citep[see, e.g.,][]{McClintockRemillard2003}.  In the
``standard'' models \citep[see][and references
therein]{ReynoldsNowak2003} the soft component is well-explained with
thermal emission from a standard thin disk \citep{ShakuraSunyaev1973},
while the hard power-law component is generally attributed to inverse
Compton (IC) scattering processes.  The various models currently in
existence often have quite different seed photons and system
geometries, yet predict similar results for the broad continuum
emission \citep[see][]{NowakWilmsDove2002,Markoffetal2003}.  In order
to discern between the models, therefore, one has to look at finer
details which are dependent upon specific elements of the geometry.

In the hard state, the power-law component dominates the thermal disk
emission over most of the X-ray range.  The presence of a cold, thin
disk can still be inferred via detection of a soft component in the
$\approx$0.3--1\,keV band, and via spectral components in the
$\approx$2--50\,keV band  suggesting that a fraction of the hard
X-rays is reprocessed or reflected from an optically thick surface.
The reflection component is characterized by a flattening of the
power-law above $\sim10$\,keV
\citep[e.g.,][]{Poundsetal1990,GeorgeFabian1991}, as well as by
spectral features such as an Fe K$\alpha$ fluorescent line and an Fe
edge \citep[see][for a review]{ReynoldsNowak2003}.  The strength of
these components is directly related to the spectrum and flux hitting
the disk, and is therefore sensitive to assumptions about the system
geometry.  For example, in models where the hard X-rays are due to
IC in a hot coronal plasma completely ``sandwiching'' the disk,
the reflection is easily too high for typical X-ray binary spectra
(and the self-consistently derived coronal temperatures are too low;
e.g., \citealt{stern:95a,dove:97a}).  Thus modifications have been
proposed, such as a recessed thin disk \citep{dove:97b,poutanen:97b},
beaming away from the disk \citep{reynolds:97c,Beloborodov1999}, or
large amounts of ionization of the disk
\citep{ross:99a,nayakshin:01a,done:01c}.

In addition to the standard corona models,
\citet{MarkoffFalckeFender2001} proposed that the entire broadband
spectrum of hard state XRBs could instead result from synchrotron
emission at the beginning of an acceleration region.  While
controversial, this model succeeds at explaining the tight correlation
of radio and X-ray emission seen in several sources \cite[e.g.,
GX~339$-$4,][]{Corbeletal2000,Corbeletal2003,Markoffetal2003}, which,
in fact, may be a universal correlation in XRB hard states
\citep{GalloFenderPooley2003}.  It is also the first model to provide
a link between the inferred presence of a hot, magnetized electron
plasma near the inner regions of the central engine to the hot,
magnetized plasma that we know exists via imaged radio jets
\citep[e.g.,][]{Stirlingetal2001}.  A shortcoming of these models,
however, is that they have not yet attempted to explain detailed X-ray
spectral features such as reflection and iron lines.  In this Letter
we calculate the expected reflection fraction from several jet models
with parameters that provide good descriptions of broad band features
(e.g., overall luminosity, flat radio spectrum, X-ray spectral slope
and cutoff) for `typical' XRB data sets, such as those shown here for
the Galactic source GX~339$-$4. We then discuss the ensuing
constraints on jet models.

\section{Model Background}

The amount of X-rays from a jet falling on an element of the disk
depends upon the distance between the jet emission zone and disk
element, the angle between the jet bulk flow and the line-of-sight to
the disk, and the bulk velocity, $\beta_{\rm j}$, of the jet at that
zone.  A moderately relativistic flow is not necessarily prohibitive,
and in fact the resulting aberration has been invoked as a solution to
reduce the reflection fraction in hard state sources.
\citet{Beloborodov1999} originally proposed this scenario for Cyg X-1,
and found that they could explain the observed reflection fraction if
a dynamic corona is beamed perpendicularly away from the disk with
$\beta\sim0.3$.  \citet{MalzacBeloborodovPoutanen2001} used a more
sophisticated approach to find that the correlations between
reflection fraction and spectral index can be explained by varying the
bulk flow velocity from $\beta\sim0.3-0.7$.  These bulk velocities are
typical for weak jets; therefore, a careful treatment of reflection
from jet emission may shed light on how models of dynamic coronae can
be unified with models of the jet base.

\clearpage

\begin{figure}
\epsscale{1}
\plotone{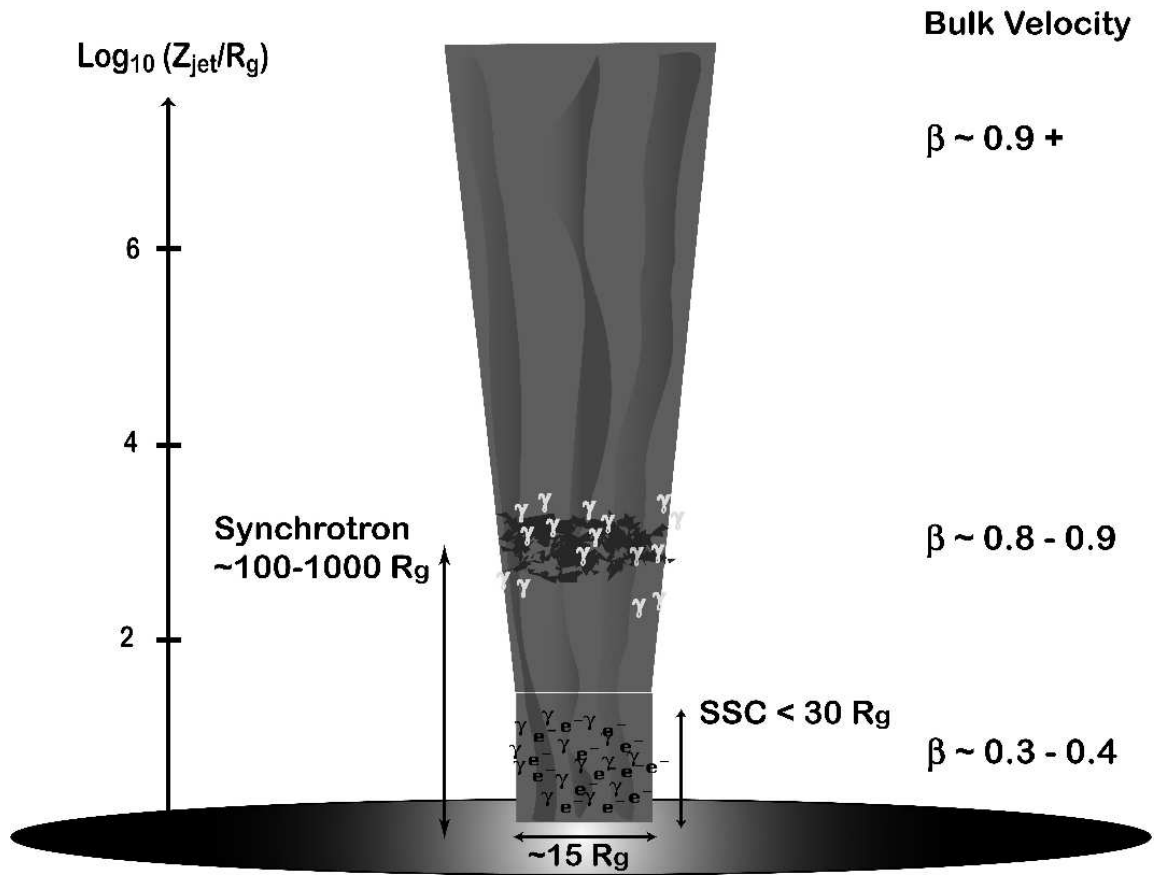}
\caption{Schematic of jet models.\label{schem}}
\end{figure}  

\clearpage

Following \citet{Beloborodov1999}, the ``reflection fraction'' of a
relativistically beamed corona can be defined as
\begin{equation}
R(\mu) \equiv { \Big \langle \frac{dP}{d\Omega} \Big \rangle} \Big /
              { \frac{dP(\mu)}{d\Omega} } ~~,
 \label{eq:def}
\end{equation}
where $dP/d\Omega$ is the emitted power per unit solid angle, $\mu
\equiv \cos \theta$, $\theta$ is the angle between the direction of
coronal bulk motion and the observer's line of sight, and the angular
average is over the half of the sky subtended by the disk.  In the
absence of beaming, $R$ would be independent of angle and $\approx 1$,
since the disk subtends nearly half the sky as viewed from the dynamic
corona.  The same would be true for our jet models, as the distance
from the jet base to its outer X-ray emission region is substantially
smaller than the outer disk radius. Beaming serves to reduce the
reflection fraction by enhancing the coronal or jet emission along
lines of sight in the direction of motion, while simultaneously
reducing the amount of emission towards the disk. 

In the specific model considered by \citet{Beloborodov1999}, the ratio
of these two effects yields a reflection fraction of
\begin{equation}
 R = { \left( 1 - \beta/2 \right ) \left ( 1 - \beta \mu \right
 )^3 } { \left ( 1 + \beta \right ) ^{-2} } ~~.
 \label{eq:frac}
\end{equation}
Even for $\beta \approx 1$, one expects a reflection fraction of 3\%
for $\mu = 1/2$.  Note that a simple answer is obtained as
\citet{Beloborodov1999} considers a photon flux per unit energy
$\propto E^{-2}$.  The spectrum shape, although not its amplitude, is
independent of angle, allowing the definition given above.

For the case of our jet models, the spectrum intercepted by the disk
has a similar, but not identical, shape to that of the directly viewed
spectrum.  This makes a translation into a simple value for
`reflection fraction' less straightforward, although one still expects
fitted values to yield $R \approx 3\%$ if $\beta \approx 1$.  Larger
reflection fractions can be achieved by decreasing the jet speed and
by increasing the fraction of X-ray emission occurring close to the
disk surface, e.g., by Compton emission from the jet base.

In the jet models considered to date, there have been propositions for
both synchrotron \citep{MarkoffFalckeFender2001,Markoffetal2003} and
external IC (EIC) \citep{Georganopoulosetal2002} contributions to the
X-rays.  In the former models, the components from synchrotron
self-Compton (SSC) and EIC from the disk were calculated
self-consistently, but were found to fall below the synchrotron
emission for the jet parameters considered.  The relative Compton
contributions can be raised if the assumed scale for the jet is
expanded, resulting in lower densities and thus allowing higher
electron temperatures.  In the synchrotron dominant models that we
have previously studied, we had assumed the radius of the jet base to
be on the order of the event horizon, specifically, we set $r_0\sim3
r_{\rm g} \equiv 3~GM/c^2$ \citep{Markoffetal2003}.  If, on the other
hand, the jet base is contiguous with, or generated in, an extended
corona, a larger scale may be more sensible.  With this in mind, we
have explored a new range of models with $r_0\sim15-20r_{\rm g}$.

The dependence of the calculated spectrum upon the model parameters
(jet size scale, jet power, electron temperature, etc.), and the
interdependence of the model parameters given the assumptions of a
`maximal jet', are complex.  Increasing the scale of the jet base
allows one to decrease the electron density as well as the magnetic
field, for a fixed equipartition relationship.  This allows one to
consider electron temperatures up to a few times higher than those
used in our previous models (to make up for lost synchrotron flux).
The higher electron temperatures lead to greater Compton emission
relative to synchrotron processes in the X-ray band.  It is important
to note that, compared to the synchrotron emission, the Compton
emission occurs close to the central black hole in a region of lower
bulk velocities ($\beta \approx 0.3$).  However, the beaming is
enough to significantly reduce the amount of reprocessed disk
radiation reaching the jet for inverse Compton upscattering.  The
photon field from the rest frame synchrotron emission in the jet is
typically orders of magnitude higher than even the direct disk photon
field.  Reprocessed disk radiation will be significantly less and thus
its feedback on the X-ray spectrum will be negligible.  We thus do not
include this in the following calculations.

In Fig.~\ref{schem}, we show a schematic of the jet model, indicating
the approximate locations of the synchrotron and EIC/SSC emission
regions, as well as their bulk velocities.  The jet base starts out as
a nozzle flow of constant radius moving at the speed of sound,
$\beta_{\rm s}=\sqrt{(\gamma_{\rm a}-1)/(\gamma_{\rm a}+1)}\sim0.4$
for our adopted adiabatic index $\gamma_{\rm a}=4/3$.  In the case of
a free jet, which we assume here for simplicity, it accelerates only
due to its longitudinal pressure gradient.  The velocity profile of
the jet, $\beta_{\rm j} (z)$, is then determined by solving the Euler
equation along the jet axis \citep[see][]{Falcke1996}. Ignoring terms
of order $\ln \gamma_{\rm j}\beta_{\rm j}$, one finds $\gamma_{\rm
j}\beta_{\rm j}\simeq\sqrt{\beta_{\rm s}^2(\gamma_{\rm
a}+4\ln(z/z_{\rm 0}))}$, where $z_{\rm 0}$ is the length of the
nozzle.  The jet also expands laterally with $\beta_{\rm s}$, and the
resultant adiabatic cooling of the particles is taken into account.

Models and scaling arguments based on active galactic nuclei (AGN)
jets suggest that the turnover from optically thick to thin jet
synchrotron in XRBs occurs in the IR/NIR.  Such a turnover is seen
explicitly in a 1981 observation of GX~339$-$4
\citep{CorbelFender2002}.  For our model, this corresponds to a region
at $\sim$100--1000\,$r_{\rm g}$ in the jet, and represents the start of the
acceleration zones.  In contrast, both SSC and EIC from the disk are
strongest at the base of the jet in the nozzle regime.

\clearpage

\begin{figure*}
\epsscale{1}
\plottwo{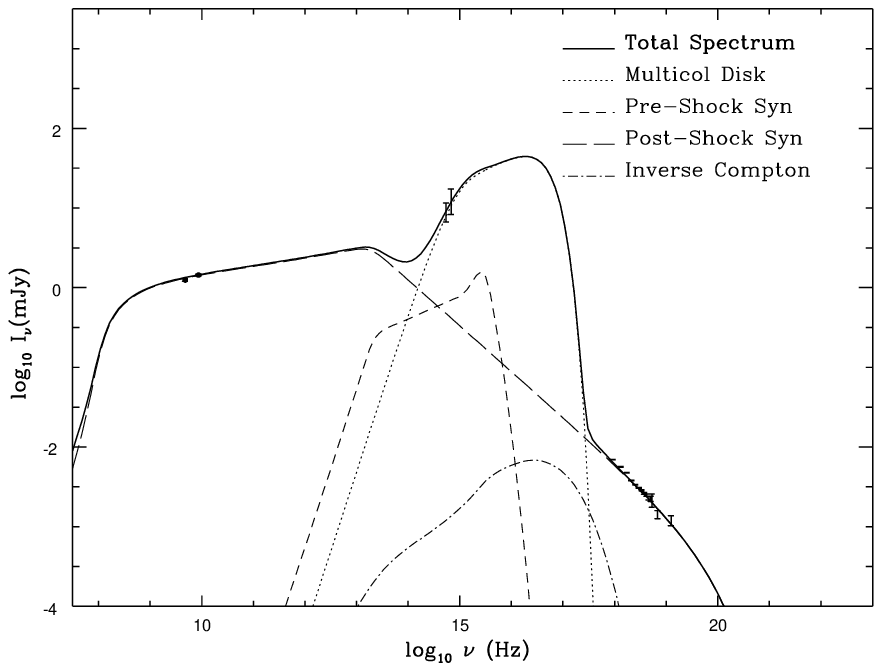}{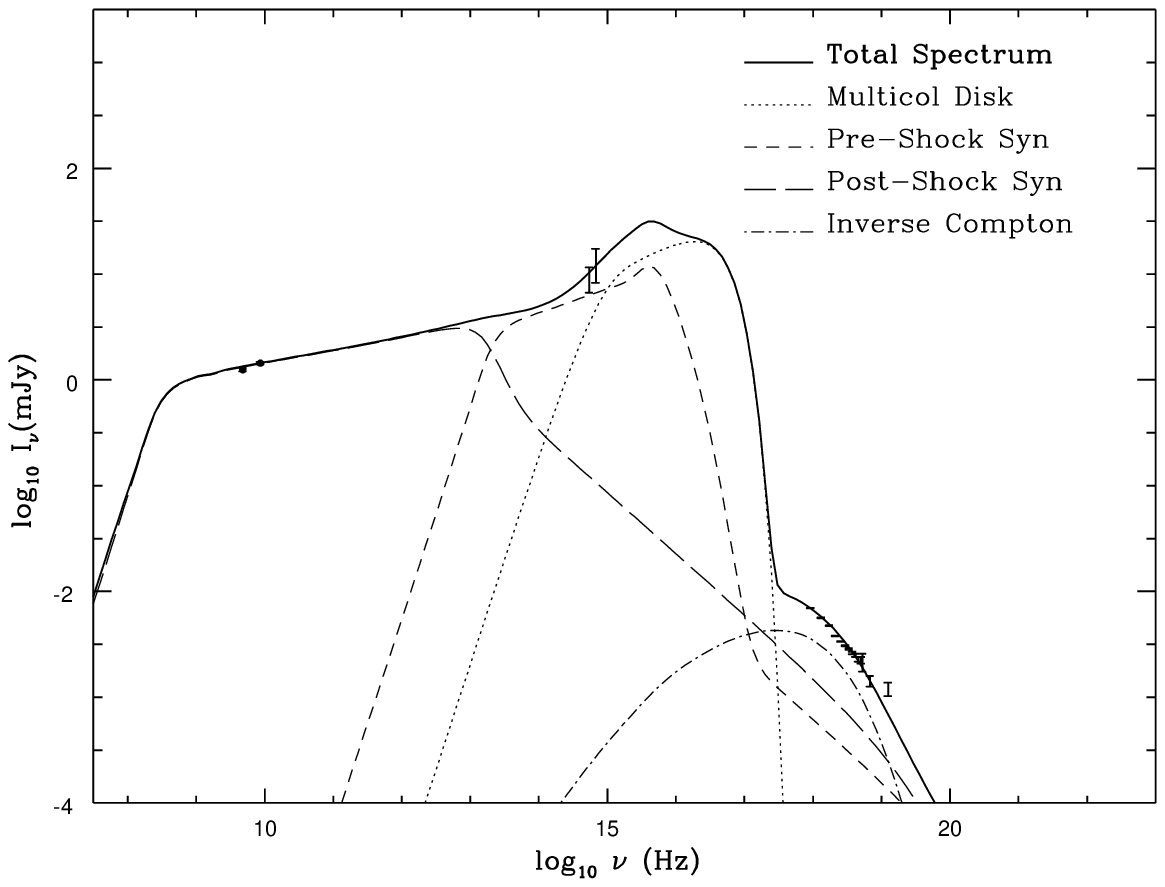}
\caption{Jet model representative fits for multiwavelength GX~339$-$4
observations from 1999 May 14 (see Nowak, Wilms, \& Dove 2002).  a)
Synchrotron-dominated regime similar to what was presented in
\citet{Markoffetal2003}, but with jet nozzle $r_0=15$ $r_{\rm g}$ and
electron temperature $T_{\rm e}=7\times10^9$ K, with equipartition parameter
$k\equiv B^2/8\pi / \int n_e(E_e)dE_e= 3$.  Roughly 10\% of the
particles are accelerated.  b) Synchrotron self-Compton
dominated regime with jet nozzle $r_0=15$ $r_{\rm g}$ and electron
temperature $T_{\rm e}=2\times10^{10}$ K, with k=30.  In this fit,
$\lesssim 1\%$ of the particles are accelerated, further suppressing
the synchrotron component.  \label{gx339}}
\end{figure*}  

\clearpage

\section{Calculation}
\subsection{Direct Jet Emission}

In order to test the amount of reflection expected from both regions
in our model, as well as to test the dependence upon the plasma
velocity, we have performed four calculations.  We have calculated the
spectra of both a synchrotron-dominated and an SSC-dominated model,
and for each type of model we employ $\beta_{\rm s} \sim0.4$, as
described above, as well as a slower jet model with $\beta_{\rm
s}\sim0.3$ (with $\beta_{\rm j}(z)$ scaled accordingly).  In
Fig.~\ref{gx339} we show representative plots of a synchrotron- and an
SSC-dominated model for one of the simultaneous radio, IR and X-ray
data sets used in \citet{Markoffetal2003}.  This particular data set
is from 1999 May 14 (RXTE observation 40108-02-03-00, labeled
``99\_3'' in Markoff et al.  2003). It represents a source hard state,
several months after the return from a prolonged soft state, as the
source was fading into quiescence.  This particular observation had a
reflection fraction of $R \approx 0.1$ when fit by coronal models (the
{\tt eqpair} code; \citealt{Coppi1999}). 

\begin{figure}
\epsscale{1}
\plotone{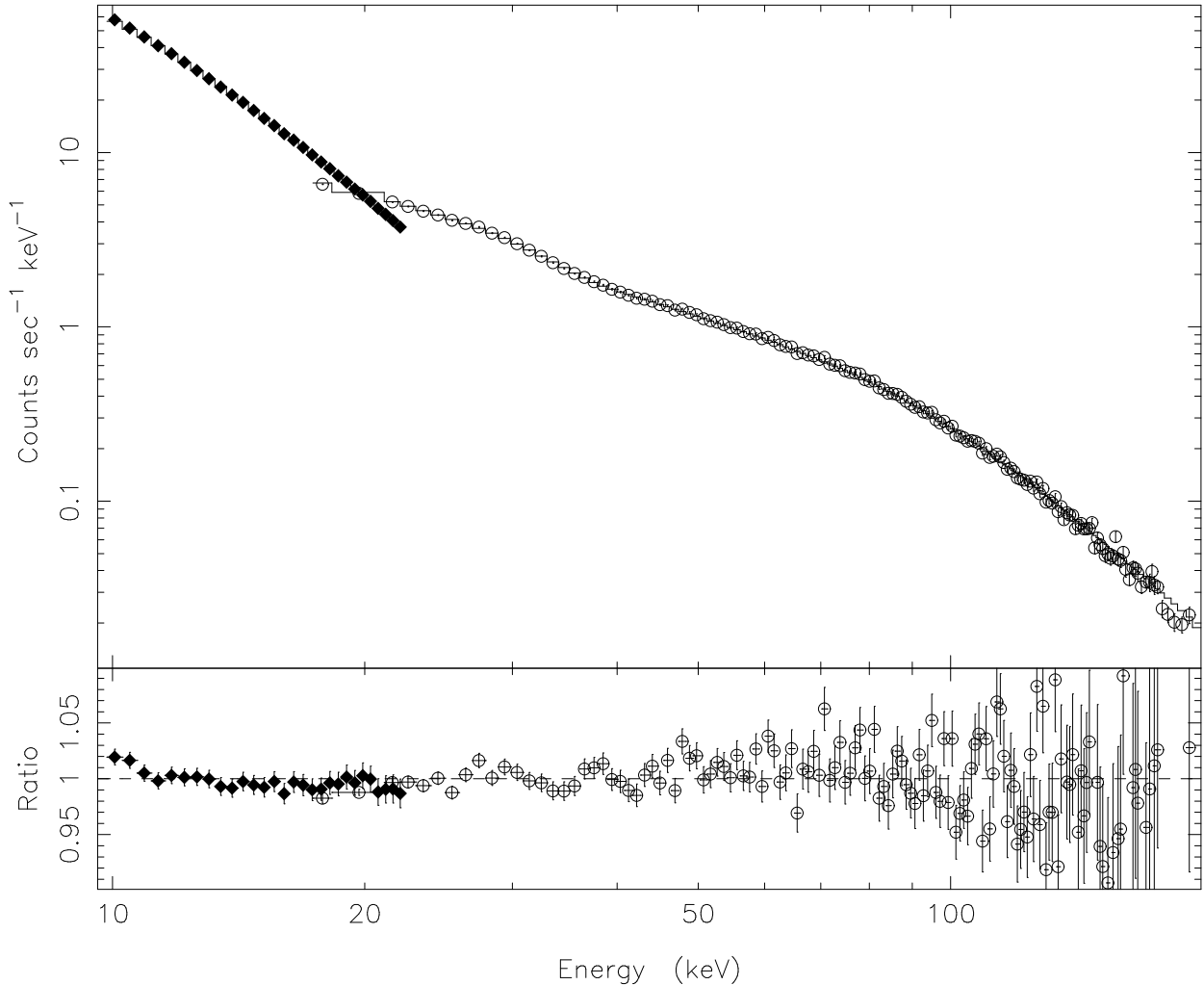}
\caption{Jet model (similar to that shown in Fig.~\ref{gx339}b),
without reflection components, fit to 10-200 keV spectra from the
low/hard state of Cyg X-1.  This spectrum has been previously
presented by \citet{Pottschmidtetal2003} (rev. 11 in that paper).  These
data are from the RXTE instruments PCA (10-22 keV; solid diamonds)
and HEXTE (18-200 keV; clear circles), extracted using the {\tt
HEASOFT v5.3} tools.  Systematic errors of 0.5\% have been applied to
the PCA data; however, no systematic errors have been applied to
HEXTE.  The PCA data has been rebinned to have a minimum of 20 counts
per bin, while HEXTE has been rebinned to have a minimum
signal-to-noise ratio of 8 in each bin.  Using {\tt ISIS v.1.1.7}
\citep{houck:00a} to perform the fits, we obtain a reduced
$\chi^2 = 1.6$ for 173 degrees of freedom.  The quality of this data
description, including the excellent fit to the high energy rollover,
is comparable to that of coronal Comptonization fits
\citep{Pottschmidtetal2003}.\label{cygx1}}
\end{figure}

The models in Fig.~\ref{gx339} have not been convolved with the
raw X-ray data through the detector response matrices.  We recently
have succeeded in importing the jet continuum model into the {\tt
XSPEC} software analysis package\footnote{The fit shown below,
however, was performed in {\tt ISIS v1.1.7} \cite{houck:00a}, which
incorporates all {\tt XSPEC} models including `local models', and will
allow us to more readily include radio and optical data in future
fits.}  \citep{arnaud:96a}, in order to begin addressing the detailed
features of the spectrum.  For instance, there has been some question
as to whether jet models can describe the shape of the spectrum near
the $\sim100$ keV cut-off region in the hard state.  We will focus on
this question in detail elsewhere (Markoff \& Nowak, in prep.);
however, here we include for reference a preliminary figure for the
Galactic BHC Cyg X-1 (Fig.~\ref{cygx1}), which exhibits evidence of a
rather steep cut-off.  Applying the jet model to the $10-200$ keV
region, we obtain a very good description of the broadband X-ray
continuum, including the turnover region.  This is a promising start,
but we need to consider further details such as line emission and a
soft component. We further need to determine how to incorporate the
reflection results presented here into a self-consistent, but also
time-efficient, fitting procedure.

\clearpage

\clearpage

\subsection{Reflected Jet Emission}

In order to calculate the reflection spectra of the jet, we divide the
jet into $\sim80$ logarithmically spaced slices along its axis, from
$ \sim 3-10^8$ $r_{\rm g}$.  The effect of relativistic
beaming is applied to the emission from each jet element, and we
calculate the angle-dependent flux intercepted by an annulus of the
disk, which is divided into 100 logarithmically spaced annuli
between radii of $6\,r_{\rm g}$ to $10^5\,r_{\rm g}$.  For a given
disk annulus, the total intercepted flux is calculated by integrating
the beamed jet emission over the full extent of the jet.  We have not
considered the effects of gravitational focusing of the emission
towards the disk; however, this is likely to be negligible for the
synchrotron-dominated jets where the X-ray emission is coming from
$\sim$100--1000\,$r_{\rm g}$.  We are likely, however, to be underestimating
the amount of reflection from the SSC-dominated jets where a
significant fraction of the emission occurs close to the inner disk.

Our approach is to calculate the emission from each segment of the jet
as if it were coming from the center of the segment along the jet
axis, which is a reasonable approximation from far away.  At its very
base, however, the jet radius is larger than the inner radius of the
disk.  In order to account for this region, we perform a separate
calculation of the nozzle emission at the jet base.  Using $\mu=-1$ in
the beaming factor, we integrate the jet base emission impinging upon
the disk out to $r_0$.  The direct emission from the jet, taking
$\mu=0.77$, is calculated as described by \citet{Markoffetal2003}.  We
further assume that observer only sees X-ray emission from one side of
the jet and disk.

The integrated jet emission hitting each radial bin is then taken and
passed through an ionized reflection code ({\tt pexriv}) and then
relativistically smeared using a Schwarzschild metric, as appropriate
for Keplerian rotation at each radius
\citep[see][]{Magdziarz1995,Zdziarskietal1995,Zyckietal1997}.  The
specific usage of these widely available codes was adapted from the
{\tt eqpair} code, where we have taken advantage of the {\tt eqpair}
spline fits of the continuum spectrum before it is passed to the
reflection routines \citep{Coppi1999}.  For Fig.~\ref{fig:ref} we have
taken a neutral disk with solar abundances, and we adopt $\mu=0.77$
for the reflected spectrum, as with the direct jet emission.

\section{Results and Discussion}

As a first step towards judging the magnitude of the reflection
fraction, we can calculate $R(\mu)$ as a function of energy by
substituting $dP/d\Omega d\nu$ into eq.~\ref{eq:def}.  The
synchrotron-dominated cases show the smallest overall reflection.  The
ratio is only $\approx$1--2\% in between energies of 1--100\,keV
($\beta_{\rm s}=0.3, ~0.4$).  The ratio peaks at 0.3\,keV with value
6\% and 8\% for $\beta_{\rm s}=0.3$ and $\beta_{\rm s} = 0.4$,
respectively.  The SSC-dominated jet shows significantly more
reflection.  For $\beta_{\rm s}=0.3$ and $\beta_{\rm s}=0.4$, the
ratio is $>10\%$ in between 0.5--23\,keV.  The former peaks at 18\% at
6\,keV, while the latter peaks at 17\% at 5\,keV.  Such values of
`reflection fraction' are comparable to the observed range for
GX~339$-$4, and in fact are larger than fitted for this particular
observation \citep{NowakWilmsDove2002}.

\clearpage

\begin{figure*}
\epsscale{1}
\plottwo{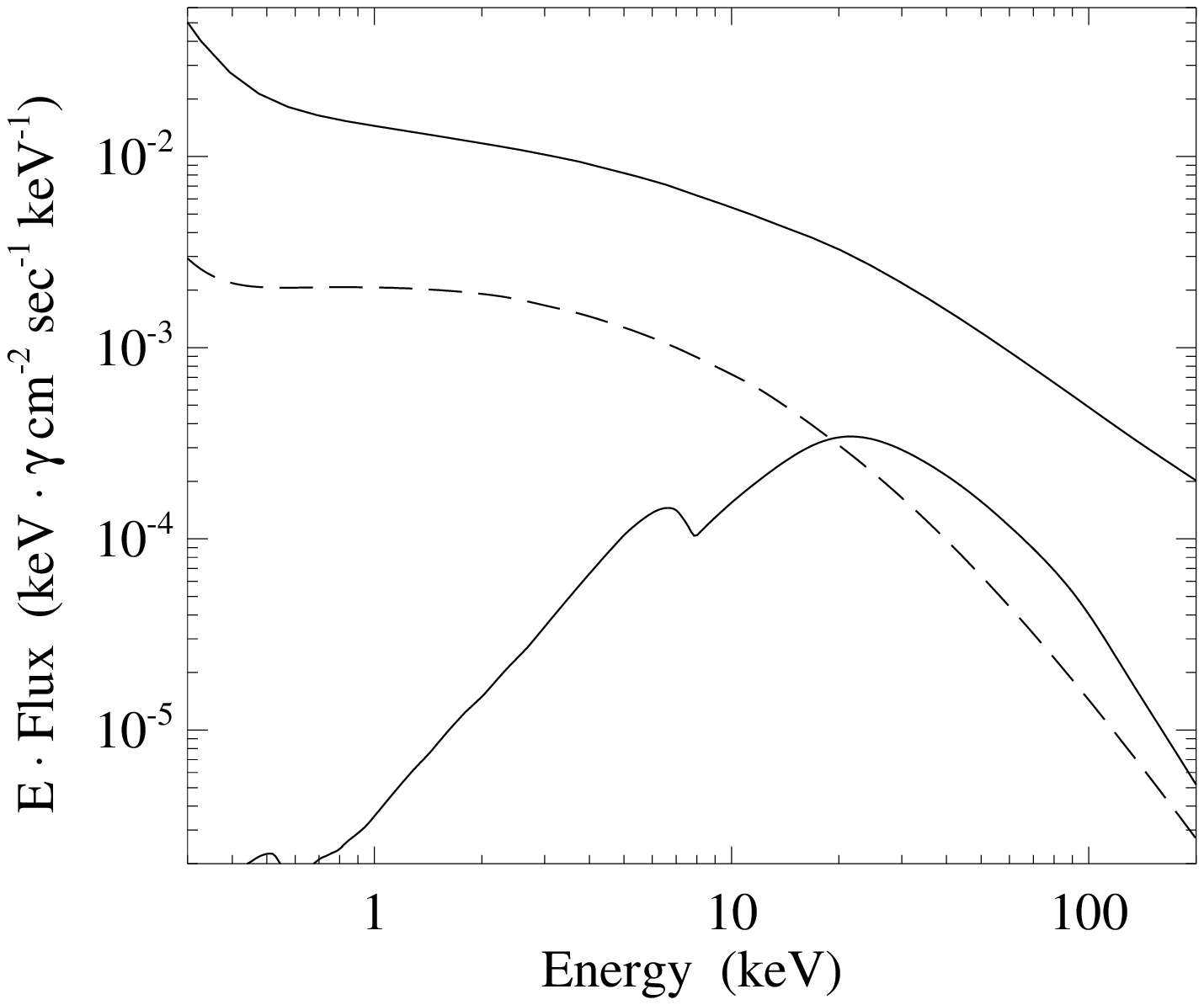}{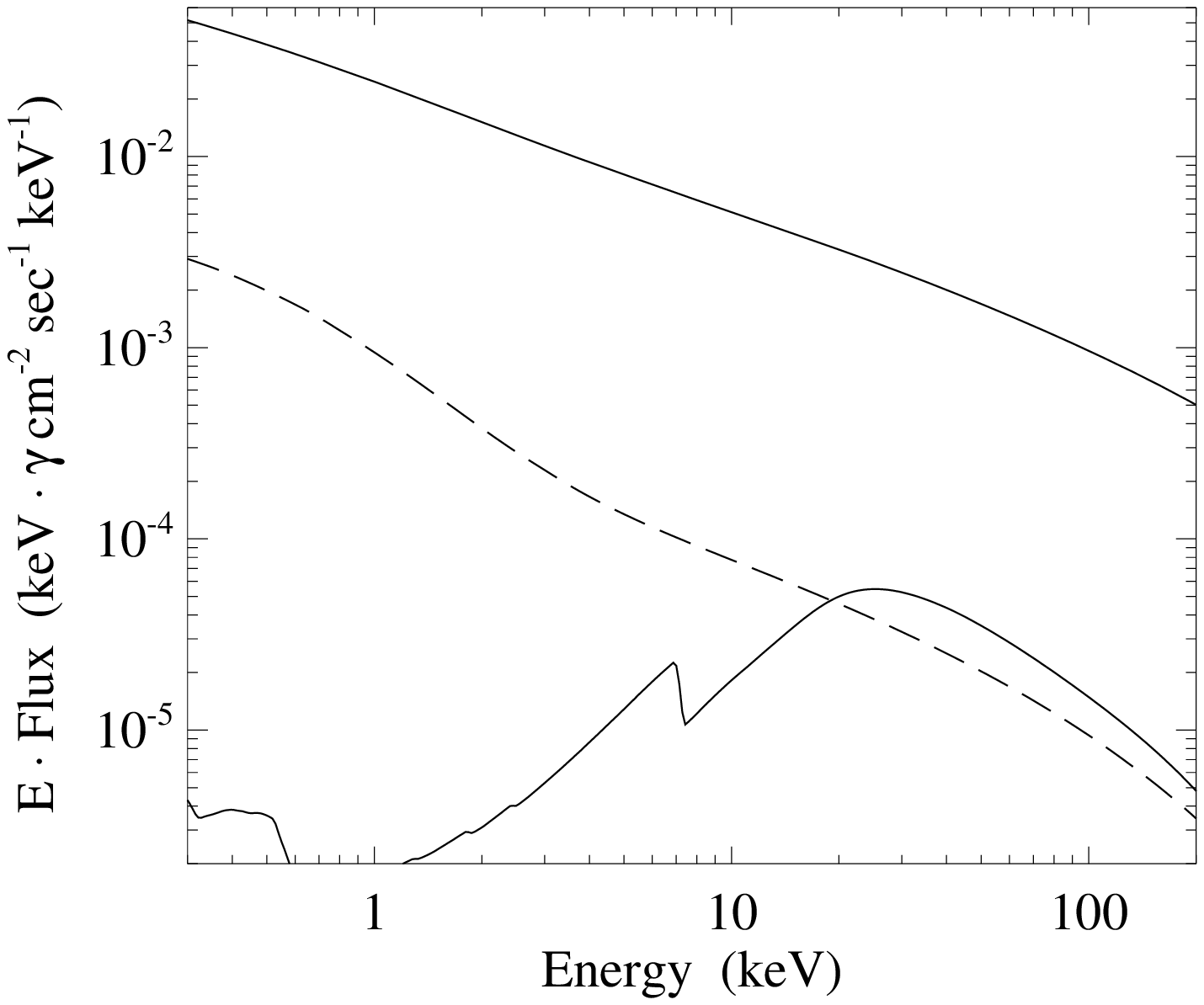}
\caption{Total model spectrum (direct plus reflected - solid line),
summed spectrum incident upon the disk (dashed line), and reflected
spectrum (lower solid line).  {\it Left:} SSC-dominated jet with
$\beta_{\rm s} = 0.4$.  {\it Right:} Synchrotron-dominated jet with
$\beta_{\rm s}
= 0.4$. \label{fig:ref}}
\end{figure*}  

\clearpage

As described above, we have also directly calculated the reflection
spectrum expected from our models.  We show spectra for the
SSC-dominated jet and the synchrotron-dominated jet, both with
$\beta_{\rm s} = 0.4$, in Fig.~\ref{fig:ref}.  Qualitative differences
are immediately apparent.  The SSC-dominated jet clearly has an
overall brighter reflection spectrum, and effects of relativistic
smearing are readily visible.  These effects are to be expected given
both the lower velocities of the SSC-dominated region ($\beta \sim
0.3$) as well as the fact that this region is closer to the central
black hole.

We have chosen a slightly face-on orientation, $\mu=0.77$, since there
has been some suggestion that this is indeed the case for the
GX~339$-$4 system \citep{Wuetal2001}.  Most BHC systems should have a
higher inclination, which, given eq.~\ref{eq:def}, allows for even
greater reflection fractions.  If we freeze all of our model
parameters but instead choose $\mu=0.5$, reflection fractions, as
defined by the ratio of the disk-incident spectrum to the directly
viewed spectrum, increase.  Over the 1--20\,keV interval, this ratio
is everywhere $> 26\%$ for the SSC-dominated jet with $\beta_{\rm s} =
0.3$, and $>7\%$ for the synchrotron-dominated jet with $\beta_{\rm s}
= 0.4$. The former peaks at 38\% at 2\,keV and, again, exhibits clear
relativistic broadening of the reflection features.

While still preliminary, this is the first time that the reflection of
jet emission off an accretion disk has been calculated for an X-ray
binary system.  These results show that jets are indeed capable of
producing the reflection fraction inferred from the X-ray data,
providing further support for a connection between the base of the jet
and the corona.  We have highlighted two extreme cases which might be
applicable to different physical situations.  Systems that exhibit
very low reflection ($\lesssim 5\%$), with sharp (i.e.,
non-relativistic) features in any reflected spectrum \citep[e.g., XTE
J1118+480,][]{Milleretal2001}, could be synchrotron-dominated, and
clearly rule out SSC-dominated jets.  Systems with significantly
larger reflection fractions ($\gtrsim 15\%$) cannot be
synchrotron-dominated, especially if they exhibit features which are
unambiguously relativistically smeared.  However even in the
Compton-dominated regime, as shown in Fig.~\ref{gx339}, synchrotron
radiation can contribute $\gtrsim 10\%$ of the flux which, as we
discuss further below, will greatly effect fits to data with corona
plus reflection models.

For intermediate values, or values $\gtrsim 30\%$, other factors need
to be considered.  For instance, these results assume that the disk is
perfectly flat, and that the jet is always perpendicular to the disk.
Realistically, disks are expected to flared or warped
\citep[e.g.,][]{Dubusetal1999}, and several systems also show evidence
for misalignment between the jets and outer disks
\citep{Maccarone2002}.  Both of these effects will serve to increase
the reflection fraction from the jet, particularly for the synchrotron
component.  Therefore, we treat these numbers as lower limits.  

We would like to emphasize, however, that a significant X-ray
contribution from jet synchrotron emission can greatly alter how one
even defines `reflection fraction' based upon a presumed
single-component, underlying continuum.  As shown in Fig.~\ref{cygx1},
jet radiation alone can provide a good description of the high
energy cutoff region.  One can readily imagine a model wherein the
soft X-ray region is dominated by SSC emission (as in
Fig~\ref{fig:ref}b) and/or Comptonization of external (disk) photons,
each with a large covering factor of the disk (i.e., essentially unity
reflection fraction, for that component alone).  The hard X-ray
radiation could then be dominated by synchrotron radiation with
inherently low reflection fraction.  The net spectrum would have an
intermediate fitted value of `reflection fraction' that does not have
a `geometric interpretation' entirely appropriate for either the soft
or hard emission components. If the broad-band X-ray continua of hard
state BHC are in fact comprised of such multiple components, as in
some of the jet models presented here, then this calls into question
current interpretations of `reflection fractions' based upon single
component fits.

Of course, in order to determine whether such multiple spectral
components are indeed present in the observations, actual fitting of
the combined direct plus reflection spectrum needs to be performed.
The calculations and models presented here provide vital clues as to
how much each process can contribute for this next step. This work has
shown that this type of analysis may hold the key to disentangling
the emission processes relevant from the accretion inflow and outflow,
and place limits on the synchrotron vs. Compton contributions to the
hard state spectrum.

\acknowledgments

We would like to thank Jon Miller for encouraging us to make this
calculation, and J\"orn Wilms for a careful reading of the manuscript.
S.M. is supported by an NSF Astronomy \& Astrophysics postdoctoral
fellowship, under NSF Award AST-0201597. This work has also been
supported by NSF Grant INT-0233441 and NASA Grant NAS8-01129.

%\bibliographystyle{apj}
%\bibliography{refs}

\end{document}